\documentclass[pre,twocolumn]{revtex4-1} 
\usepackage{graphicx}
\usepackage{verbatim}
\usepackage{caption}
\usepackage{amsmath}
\usepackage{color}
\usepackage[dvipsnames]{xcolor}
\usepackage{multirow}

\begin{document}
\newcommand{\tca}[1]{\textcolor{blue}{#1}}
\newcommand{\tcom}[1]{{\it\textcolor{magenta}{#1}}}

\title{A novel numerical scheme for nonlinear electron-plasma oscillations}

\author{Prabal Singh Verma\footnote{prabal-singh.verma@univ-amu.fr}}
\affiliation{CNRS, Aix-Marseille Univ., PIIM, UMR 7345, Marseille F-13397, France}

\date{\today}

\begin{abstract}
In this work, we suggest an easy-to-code higher-order finite volume semi-discrete scheme to analyze the nonlinear behavior of the electron-plasma oscillations by solving electron fluid equations numerically. The present method employs a fourth-order accurate centrally weighted essentially nonoscillatory reconstruction (CWENO4) polynomial for estimating the numerical flux at the grid-cell interfaces, and a fourth-order Runge-Kutta method for the time integration. The numerical implementation is validated by reproducing earlier results for both non-dissipative and dissipative cold plasmas. The stability of the present scheme is illustrated by evolving the nonlinear electron plasma oscillations in a 
cold non-dissipative plasma for hundred plasma periods, which also display a negligible numerical dissipation. The fourth-order accuracy of the existing approach is also confirmed by evaluating the convergence of errors for the nonlinear electron plasma oscillations in a cold non-dissipative plasma.

\end{abstract}


\maketitle

\section{Introduction}\label{sec:intro}
Numerical simulations have always been very useful to illustrate the physics of a strongly nonlinear system for which an analytical
approach is not feasible \citep{jain2003nonlinear,verma2010nonlinear,verma2012breaking,verma2012residual,verma2017generation,verma2017nonlinear}.
The primary focus of this manuscript is to introduce a new numerical scheme for studying nonlinear electrostatic plasma oscillations in various physical regimes. Investigation of nonlinear electrostatic oscillations in different plasmas is a fascinating field of research for more than five decades \citep{dawson1959nonlinear,davidson1968nonlinear,kaw1973quasiresonant,stenflo1998electron,gupta1999phase,rowlands2008exact,sengupta2009phase,maity2010nonlinear,
maity2012breaking,maity2013wave,brodin2014large,brodin2014nonlinear,brodin2017simple,brodin2017nonlinear}. One of the leading application of 
these studies is in the laser or particle beam induced wake-field acceleration experiments and simulations, where one aims at attaining mono-energetic charged particles in short distances, e.g. \cite{faure2004laser,jain2015positron}. Depending upon the problem of interest multiple numerical methods, for example, the Particle-In-Cell (PIC) simulation \cite{birdsall2004plasma}, the Vlasov simulation, e.g. \citep{mandal2016study,schamel2017nonlinear}, the sheet simulation \citep{sengupta2009phase,verma2012breaking} and the direct numerical simulations of the fluid equations, e.g. \citep{jain2003nonlinear,verma2010nonlinear,bera2015fluid,bera2016relativistic} are being practiced widely in the literature.
In the PIC simulation, charged particles are evolved in time by solving the coupled Lorentz force equation and Maxwell's equations. In the Vlasov simulation, the distribution function of a system, combined with the Poisson's equation, is developed in time. The sheet simulation deals with the evolution of the physical quantities in Lagrange coordinates and is quite similar to the fluid simulation in the sense that it leads to unphysical results after the sheet-crossing (wave-breaking) \cite{dawson1959nonlinear}.

Wave-breaking is a nonlinear physical process that translates the coherent energy into random kinetic energy as a result of the wave-particle interactions. It occurs when the nearby sheets or the fluid elements, taking part in the coherent oscillations, cross each other. One of the advantages of the PIC and the Vlasov simulations over fluid simulations is that the formers are capable of illustrating the physics of the wave-particle interactions, and therefore these numerical tools can be used to study the dynamics of the plasma species even beyond the
wave-breaking \citep{verma2012residual,verma2017generation}.
Nevertheless, these simulations are computationally very expensive and exhibit a significant noise, especially in spatial profiles of the charge density fluctuations. On the contrary, sheet simulations show negligible noise (numerical error), and it is also probable to make them deliver physical results even beyond the sheet crossing (wave-breaking). However, the main limitation of these simulations is that they cannot be stretched to study the dynamics of a multi-component plasma. Therefore, we here propose a simple numerical scheme to solve the electron fluid equations, which delivers comparatively more accurate result when the wave-particle interactions are not essential and can easily be extended to multi-species plasma.

Note here that usually a finite difference scheme based on flux corrected transport \cite{boris1997flux} is utilized to study nonlinear electrostatic oscillations in different physical regimes, e.g. \citep{verma2010nonlinear,bera2015fluid,bera2016relativistic}. As an alternative, we here present an easy-to-implement higher-order finite volume semi-discrete scheme, based on a fourth-order centrally weighted essentially nonoscillatory (CWENO4) reconstruction \citep{lpr7} and a fourth order Runge-Kutta method, for the same.  The CWENO approach, which is originally
based on WENO approach (see, for example, \cite{shu2009high}), is being used widely to solve multi-dimensional hyperbolic problems \citep{lpr1,lpr2,lpr3,lpr4,lpr5,lpr6,lpr7,verma2017higher}. We are, however, employing this classical approach, for the first time, to analyze nonlinear electrostatic oscillations in a cold electron-ion plasma, where the ions are considered to be infinitely massive. In order to validate the implementation, we first rediscover the earlier results for both non-dissipative and dissipative cold plasmas. 
Later we test the stability and the numerical dissipation of the numerical scheme by following the evolution of the nonlinear 
electron plasma oscillations up to hundred of plasma periods. We do not observe any visible dissipation and unphysical effect (e.g. numerical 
noise) in this test. 
We then confirm the fourth-order accuracy of the present numerical scheme by estimating the error convergence for nonlinear electron plasma oscillations.

The flow of the present manuscript is as follows. In section II, we introduce the electron fluid equations and rewrite them in the conservative form. In section III, a semidiscrete scheme is introduced to solve electron fluid equation numerically. Section IV deals with the validation and accuracy of the numerical implementation. In section V we provide the summary of the present work and discussion.

\section{Governing equations} \label{sec:fluid-equations}
The basic equations governing the dynamics of the electron fluid in a cold homogeneous,
unmagnetized, dissipative plasma, are the momentum equation, the continuity equation,
 Poisson's equation and the current equation:

\begin{equation}\label{mom}
\frac{\partial v_e}{\partial t} +  v_e\frac{\partial v_e}{\partial x} = -E + 
\frac{1}{n_e}\frac{\partial}{\partial x}\Big(\nu_e \frac{\partial v_{e}}
{\partial x}\Big) -  \eta v_e, 
\end{equation}
\begin{equation}\label{con}
 \frac{\partial n_{e}}{\partial t}+\frac{\partial(n_{e}v_{e})}{\partial x} = 0, 
\end{equation}
\begin{equation}\label{poi}
 \frac{\partial E}{\partial x} = 1 - n_e, 
\end{equation}
\begin{equation}\label{curr}
 \frac{\partial E}{\partial t} = v_e n_e.
\end{equation}
Here we assume that the spatial variations are only along the $x$-direction, therefore derivatives along the other directions $(y,z)$ are 
not present in the above equations. Moreover, for simplicity ions are considered to be static and homogeneously distributed in space. 
Note that the
Eqs.\eqref{mom}-\eqref{curr} are in
normalized form, where
$ x \rightarrow k^{-1}, \hspace{0.2cm} t \rightarrow \omega_{pe}^{-1} , 
 \hspace{0.2cm} n_e \rightarrow n_0, \hspace{0.2cm} E \rightarrow \frac{m\omega_{pe}^2}{ke}, 
 \hspace{0.2cm} v_e \rightarrow \frac{\omega_{pe}} {k}, 
\hspace{0.2cm} \nu_e \rightarrow \frac {mn_0\omega_{pe}}{k^2},
\hspace{0.2cm} \eta \rightarrow \frac {m\omega_{pe}}{n_0 e^2}$.
Here $\nu_e$ is the electron viscosity, $\eta$ is the plasma resistivity and rest of the symbols have their usual
definitions. Since for realistic situations $\nu_e(x,t)$ can be considered to be constant, we assume
 $\nu_{e} = \nu_0$ which now can be taken out of the partial derivative in Eq.\eqref{mom}
\cite{verma2010nonlinear}.
Thus, Eqs.\eqref{mom}-\eqref{curr} can also be expressed as,

\begin{equation}\label{mom1}
 n_e\frac{\partial v_e}{\partial t} +  n_e v_e\frac{\partial v_e}{\partial x} = -n_e E + \nu_0 
\frac{\partial^2 v_{e}}{\partial x^2} -  \eta n_e v_e, 
\end{equation}
\begin{equation}\label{con1}
 \frac{\partial n_{e}}{\partial t}+\frac{\partial(n_{e}v_{e})}{\partial x} = 0, 
\end{equation}
\begin{equation}\label{poi_curr1}
 \frac{\partial E}{\partial t}+v_{e}\frac{\partial E}{\partial x} = v_e. 
\end{equation}
Equations \eqref{mom1}-\eqref{poi_curr1} can be further written in the conservative form as follows,

\begin{equation}\label{mom2}
 \frac{\partial (n_e v_e)}{\partial t} +  \frac{\partial (n_e v_e^2)}{\partial x} = -n_e E + \nu_0 
\frac{\partial^2 v_{e}}{\partial x^2} -  \eta n_e v_e, 
\end{equation}
\begin{equation}\label{con2}
 \frac{\partial n_{e}}{\partial t}+\frac{\partial(n_{e}v_{e})}{\partial x} = 0, 
\end{equation}
\begin{equation}\label{poi_curr2}
 \frac{\partial n_e E}{\partial t}+\frac{\partial(n_e E v_e)}{\partial x} = n_e v_e. 
\end{equation}
The above-mentioned equations can be written in a more compact form:

 \begin{equation}\label{conservative}
 \frac{\partial \bf U}{\partial t}+\frac{\partial \bf F}{\partial x} = \bf S,
 \end{equation}
where ${\bf U} = (n_e v_e,n_e,n_e E)$ is the vector of conserved quantities, ${\bf F} = (n_e v_e^2,n_e v_e,n_e E v_e)$ is
the corresponding vector of fluxes along the $x$-direction, and ${\bf S} = (-n_e E + \nu_0 
\frac{\partial^2 v_{e}}{\partial x^2} -  \eta n_e v_e, 0, n_e v_e)$ is the vector of the respective source terms.

In Sec.\ref{sec:num} we will discuss the finite volume method to solve Eq.\eqref{conservative} numerically.

\section{Numerical Scheme}\label{sec:num} In order to solve Eq.\eqref{conservative} numerically we discretize the computational domain $[0, l_x]$ into small grid cells where $l_x$ is the length of the domain along the $x$-direction. Suppose $n_x$ is the number of grid cells along the $x$-direction then the grid-cell size becomes $\Delta x = l_x/n_x$. Now consider a grid cell $i$ centered at $x_i$ and perform the line integration of Eq.\eqref{conservative} about the
grid cell as,

 \begin{equation}\label{conservation1}
 \frac{1}{\Delta x} \int_{x_{i-1/2}}^{x_{i+1/2}}\mathrm{d}x  \,  \Big[ \frac{\partial \bf U}{\partial t}+\frac{\partial \bf F}{\partial x}=  {\bf S} \Big],
 \end{equation}
here $x_{i\pm1/2} = x_i \pm \Delta x/2$ correspond to the positions of
grid cell interfaces along the $x$-direction. Eq.\eqref{conservation1} can be further rewritten as,
 \begin{equation}\label{semi-dis}
 \frac{d\overline {\bf U}_{i}}{d t} = -\frac{{\bf F}_{i+1/2}-{\bf F}_{i-1/2}}{\Delta x}-\overline {\bf S}_{i}. 
 \end{equation}
Here $\overline {\bf U}_{i}$, $\overline {\bf S}_{i}$ are the averages of ${\bf U}$, ${\bf S}$, respectively in the
grid cell $i$ and are defined as,
 \begin{equation}\label{u_avg}
 \overline {\bf U}_{i}(t) = \frac{1}{\Delta x}  \int_{x_{i-1/2}}^{x_{i+1/2}}\mathrm{d}x  \, {\bf U}(x,t) \, ,   
 \end{equation}
 \begin{equation}\label{s_avg}
 \overline {\bf S}_{i}(t) = \frac{1}{\Delta x}  \int_{x_{i-1/2}}^{x_{i+1/2}}\mathrm{d}x  \, {\bf S}[{\bf U}(x,t)] \, ,   
 \end{equation}
and ${\bf F}_{i\pm 1/2}$ are the fluxes of ${\bf U}$ at the
grid cell interfaces along the $x$-direction and are defined as,
 \begin{equation}\label{F_avg}
  {\bf F}_{i\pm 1/2}(t) =  {\bf F}({\bf U}(x_{i\pm 1/2},t)),   
 \end{equation}

There are several methods to estimate the physical fluxes at the interfaces by employing {various Riemann-solvers}.
In this work we employ a
simple approximation to the Riemann problem {\it i.e.} the local Lax-Friedrichs flux (LLF), e.g.
\citep{verma2017higher}. The LLF approximation to the point-value flux at the center of  a cell interface $x_{i+1/2}$ is given by,
 \begin{equation}\label{Fx}
 {\bf F}_{i+1/2} =  \frac{{\bf F}({\bf U}_{i+1}^-)
+{\bf F}({\bf U}_{i}^+)}{2} -  
                              \frac{a_{i+1/2}}{2}  ({\bf U}^{-}_{i+1}-{\bf U}^{+}_{i}), 
 \end{equation}
where the quantities ${\bf U}_{i}^+$, ${\bf U}_{i+1}^-$ are fourth-order accurate point values of the conserved quantities at the grid cell interface $x_{i+1/2}$. They are computed from the CWENO4 polynomial ${\bf R^i}_{i}(x)$, ${\bf R^{i+1}}_{i+1}(x)$ (discussed below in the subsection \ref{sec:reconstruction}),
respectively and {${\bf F}({\bf U}_{i}^+)$, ${\bf F}({\bf U}_{i+1}^-)$} are the respective point value fluxes. The quantity  $a$
is the local maximum speed of propagation
which is estimated as (see for example, \cite{kurganov-2000}),
 \begin{equation}\label{ax}
 a^{x}_{i+1/2} = \text{max}\left\{ \rho \left(\frac{\partial {\bf F^x}({\bf U}_{i}^+)}{\partial U}\right), \\ 
                             \rho \left(\frac{\partial {\bf F^x}({\bf U}_{i+1}^-)}{\partial U}\right) \right\}\,,
 \end{equation}
where $\rho$(A) is the maximum of the magnitude of the eigenvalues of the Jacobian matrix A.

\subsection{Reconstruction of the CWENO4 polynomial from the averages} \label{sec:reconstruction} We construct here a 1D fourth-order CWENO polynomial from the given averages. Before going into the details of the procedure,
we note here that the method for reconstructing such polynomials is well described in \citep{lpr3}. Nonetheless, we provide here a summary of the method for the {sake} of easy implementation.

In this approach we reconstruct a quadratic polynomial ${\bf R^i}_{i}(x)$ in each cell $[i]$, which is a convex combination of three quadratic polynomials ${\bf P^i}_{i-1}(x)$, ${\bf P^i}_{i}(x)$ and ${\bf P^i}_{i+1}(x)$ such that,
\begin{equation} \label{R_I_4th}
{\bf R^i}_{i}(x) = \sum_{l = i-1}^{i+1} w^i_{l} {\bf P^i}_{l}(x),
\end{equation}
where the superscript `$i$' appears to distinguish the reconstruction
polynomials $\bf R$ in different grid-cells. The quantities $w^i_{l}$ are the nonlinear weights which ensure higher-order accuracy in the smooth regions and non-oscillatory
behavior near a discontinuity. These weights satisfy the following criteria,
\begin{equation} \label{weights_sum}
\sum_{l = i-1}^{i+1} w^i_{l} = 1
 ,\hspace{0.2cm} w^i_{l} \geq 0, \hspace{0.2cm}  \forall \hspace{0.1cm} l \in (i-1, i, i+1),
\end{equation}
and are defined as,
\begin{equation} \label{w_l}
w^i_{l} = \frac{\alpha^i_{l}}{\alpha^i_{i-1}+\alpha^i_{i}+\alpha^i_{i+1}},
\end{equation}
with,
\begin{equation} \label{alpha}
\alpha^i_{l} = \frac{c_{l}}{(\epsilon + IS^i_{l})^p}, \hspace{0.2cm} \forall \hspace{0.1cm} l \in (i-1,i,i+1).
\end{equation}
Here $\epsilon$, $p$ are chosen to be $10^{-6}$ and $2$, respectively and {the constants $c_{i-1} = c_{i+1} = 1/6$, $c_{i} = 2/3$
are chosen so as to guarantee the fourth order accuracy of the physical quantities at the cell-boundaries \cite{lpr3}}.
The quantity $IS^i_{l}$ is the smoothness indicator quantifying the smoothness of the corresponding
polynomial ${\bf P^i}_{l}(x)$. It is defined as,
 \begin{eqnarray}\label{ISn}
\nonumber  IS^i_{l} &=& \sum_{n=1}^2 \int_{x_{i-1/2}}^{x_{i+1/2}} \! (\Delta x)^{2n-1}  ({\bf P^i}_{l}^{(n)}(x))^2 \,  \mathrm{d}x , 
 \end{eqnarray}
where $l \in (i-1,i,i+1)$ and $n$ represents the order of the derivative w.r.t. $x$.
All the three coefficients of the quadratic polynomial ${\bf P^i}_{l}(x)$ are obtained uniquely by
imposing the conservation of the three averages,
$\overline {\bf U}_{l-1}$, $\overline {\bf U}_{l}$ and $\overline {\bf U}_{l+1}$, where $l \in (i-1,i,i+1)$.
Thus, each polynomial,
${\bf P^i}_{l}(x)$, can be written as,
\begin{eqnarray} \label{P_m}
\nonumber {\bf P^i}_{l}(x) &=& \overline {\bf U}_{l} - \frac{1}{24}(\overline {\bf U}_{l+1}-2\overline {\bf U}_{l}+\overline {\bf U}_{l-1}) 
 \\ \nonumber &&+ \frac{\overline {\bf U}_{l+1} -\overline {\bf U}_{l-1}}{2 \Delta x}(x-x_l)  
 \\   &&+ \frac{(\overline {\bf U}_{l+1}-2\overline {\bf U}_{l}+\overline {\bf U}_{l-1})} {2\Delta x^2} (x-x_l)^2, 
\end{eqnarray}
where  $l \in (i-1,i,i+1)$.
After we reconstruct all the polynomials (${\bf P^i}_{i-1}$, ${\bf P^i}_{i}$, ${\bf P^i}_{i+1}$), smoothness indicators
can easily be computed using Eq.\eqref{ISn}. Nevertheless, we provide here the final expressions of the same for an easy reference.
 \begin{eqnarray}\label{ISn_cweno4}
 \nonumber IS^i_{i-1} &=& \frac{13}{12} (\overline {\bf U}_{i-2} - 2 \overline {\bf U}_{i-1} 
+ \overline {\bf U}_{i})^2 \\ && + \frac{1}{4} (\overline {\bf U}_{i-2}-4\overline {\bf U}_{i-1}+3\overline {\bf U}_{i})^2, \\ 
 \nonumber IS^i_{i} &=& \frac{13}{12} (\overline {\bf U}_{i-1} - 2 \overline {\bf U}_{i} 
+ \overline {\bf U}_{i+1})^2 \\ && + \frac{1}{4} (\overline {\bf U}_{i-1}-\overline {\bf U}_{i+1})^2, \\ 
 \nonumber IS^i_{i+1} &=& \frac{13}{12} (\overline {\bf U}_{i} - 2 \overline {\bf U}_{i+1} 
+ \overline {\bf U}_{i+2})^2 \\ && + \frac{1}{4} (3\overline {\bf U}_{i}-4\overline {\bf U}_{i+1}+\overline {\bf U}_{i+2})^2,  
 \end{eqnarray}
Employing the values of the smoothness indicators in Eqs. \eqref{w_l}-\eqref{alpha}, nonlinear weights can be computed.
These weights are finally used to reconstruct the polynomial ${\bf R^i}_{i}(x)$ in Eq.\eqref{R_I_4th}.

Once we know the polynomial ${\bf R^i}_{i}(x)$, we can compute the values of the physical quantities at the grid-cell interface
by estimating ${\bf R^i}_{i}(x\pm1/2) = {\bf U}^{\pm}_{i}$.
For the sake of convenience,
we provide here the formula, all derivation done, for ${\bf U}^{\pm}_{i}$ from the knowledge of the surrounding cell averages:
 \begin{eqnarray}\label{U+-}
\nonumber {\bf U}_{i}^+ &=& \frac{1}{6} \Big [ w^i_{i-1}(11\overline {\bf U}_{i}-7\overline {\bf U}_{i-1}+2\overline {\bf U}_{i-2})\\
\nonumber &&+w^i_{i}(2\overline {\bf U}_{i+1}+5\overline {\bf U}_{i}-\overline {\bf U}_{i-1})\\
&&+w^i_{i+1}(2\overline {\bf U}_{i}+5\overline {\bf U}_{i+1}-\overline {\bf U}_{i+2}) \Big ], \\ 
 \nonumber {\bf U}_{i}^- &=&\frac{1}{6}\Big [ w^i_{i-1}(2\overline {\bf U}_{i}+5\overline {\bf U}_{i-1}-\overline {\bf U}_{i-2})\\
 \nonumber &&+w^i_{i}(2\overline {\bf U}_{i-1}+5\overline {\bf U}_{i}-\overline {\bf U}_{i+1})\\
&&+w^i_{i+1}(2\overline {\bf U}_{i+2}-7\overline {\bf U}_{i+1}+11\overline {\bf U}_{i}) \Big ].
 \end{eqnarray}
The values of ${\bf U}^{\pm}_{i}$ allow us to compute the fluxes ${\bf F}_{i+1/2}$ from Eq.\eqref{Fx}.
In the subsection \ref{sec:source} we would discuss how to deal with the source terms.

\subsection{Source terms}\label{sec:source} We need to be cautious when computing the source terms especially when they are not conservative variables. For example, the viscous term contains a second-order derivative of the velocity $v_e$ (not $n_e v_e$). Note that $v_e$ is not a conservative variable, therefore, in order to maintain the fourth-order accuracy we first have to obtain point values for the density $n_e$ and the momentum density $n_e v_e$. Then we need to divide the point-value of the momentum density by the point-value of the electron density $n_e$ to compute fourth-order accurate point-value of the velocity $v_e$. For a given averaged quantity $\overline{\bf U}$, a fourth-order accurate point-value ${{\bf U}}_{i}$ of at the cell center can be obtained as follows:

\begin{eqnarray}\label{ue_2d}
 {{\bf U}}_{i} =   \overline{\bf U}_{i} 
                                           -\frac{1}{24} ( \overline{\bf U}_{i-1}  
                                - 2 \overline{\bf U}_{i} + \overline{\bf U}_{i+1}) + {\bf\mathcal O}(\Delta x^4 ). 
 \end{eqnarray}

Once we know the point values, a fourth-order approximation to the second-order derivative w.r.t. $x$ can be obtain
as below,
 \begin{eqnarray}\label{S2}
\nonumber  {U_i}_{xx} &=& \frac{- U_{i+2} + 16 U_{i+1} -30 U_{i} + 16 U_{i-1} - U_{i-2}}{12 \Delta x^2} \\ && 
               + {\bf\mathcal O}(\Delta x^4 ) = {W_i} (say), 
 \end{eqnarray}

We then perform a fourth-order accurate averaging procedure to compute corresponding average,

\begin{eqnarray}\label{w}
\nonumber {\overline{\bf W}}_{i} &=&   {\bf W}_{i} 
                                           +\frac{1}{24} ( {\bf W}_{i-1}  
                                - 2 {\bf W}_{i} + {\bf W}_{i+1}) + {\bf\mathcal O}(\Delta x^4 ). \\ & 
 \end{eqnarray}

Note that the operations in Eqs. \eqref{ue_2d} and \eqref{w} are essential only for the source terms which are not conservative variable ${{\bf U}}$.  In case, we fail to do that the overall accuracy of the scheme will be only second-order (see Ref.\cite{verma2017higher} for a detailed description).
In the next section \ref{sec:integration}, we will discuss the time integration of the semi-discrete scheme (Eq.\eqref{semi-dis}).

\subsection{Time integration of the semidiscrete scheme} \label{sec:integration} {After computing the fluxes ${\bf F_{i\pm1/2}}$ and the source terms $\overline{\bf S_i}$, Eq.\eqref{semi-dis} is evolved using a classical fourth order accurate low-storage Runge-Kutta method \cite{rk4} to achieve the fourth-order accuracy during the temporal evolution. The steps for the same are explained as follows: let us assume that the flux term of Eq.\eqref{semi-dis} is ${\bf C[\overline {U}_{i}]}$, now dropping the subscript `$i$' Eq.\eqref{semi-dis} can be rewritten as,
 \begin{equation}\label{semi-dis-new}
 \frac{d\overline {\bf U}(t)}{d t} = {\bf C}[{\bf \overline {U}}(t)] + \overline {\bf S}(t)   
 \end{equation}
The intermediate steps to solve Eq.\eqref{semi-dis-new} are as follows,
 \begin{eqnarray}\label{K1}
  \nonumber    {\bf K}_1 = {\bf C}[{\bf \overline {U}}(t_n)]+{\bf S}[{\bf \overline {U}}(t_n)], 
 \end{eqnarray}
 \begin{eqnarray}\label{U1}
   \nonumber   \overline{\bf U}_1 =  {\bf \overline {U}}(t_n) + \frac{\Delta t}{2} {\bf K}_1,
 \end{eqnarray}
 \begin{eqnarray}\label{K2}
   \nonumber   {\bf K}_2 = {\bf C}[{\bf \overline {U}}_1]+{\bf S}[{\bf \overline {U}_1}], 
 \end{eqnarray}
 \begin{eqnarray}\label{U2}
  \nonumber    \overline{\bf U}_2 =  {\bf \overline {U}}(t_n) + \frac{\Delta t}{2} {\bf K}_2,
 \end{eqnarray}
  \begin{eqnarray}\label{K3}
   \nonumber   {\bf K}_3 = {\bf C}[{\bf \overline {U}}_2]+{\bf S}[{\bf \overline {U}_2}], 
 \end{eqnarray}
 \begin{eqnarray}\label{U3}
   \nonumber   \overline{\bf U}_3 =  {\bf \overline {U}}(t_n) + {\Delta t} {\bf K}_3,
 \end{eqnarray}
  \begin{eqnarray}\label{K4}
   \nonumber   {\bf K}_4 = {\bf C}[{\bf \overline {U}}_3]+{\bf S}[{\bf \overline {U}_3}], 
 \end{eqnarray}
 \begin{eqnarray}\label{Un+1}
  \nonumber    \overline{\bf U}(t_{n+1}) =  {\bf \overline {U}}(t_n) + \frac{\Delta t}{6} 
({\bf K}_1+ 2 {\bf K}_2+2 {\bf K}_3+{\bf K}_4).
 \end{eqnarray}
Here, $n = 0, 1, 2, 3,....$ and ${\Delta t}$ is determined dynamically according to the
Courant-Friedrichs-Lewy (CFL) constraint (see for example, \cite{verma2017higher}),
  \begin{eqnarray}\label{dt}
      \Delta t  = C_{CFL}  min\left(\frac{\Delta x}{a_{max}}\right), 
 \end{eqnarray}
where, $C_{CFL}$ is the CFL number which for all the tests is chosen as, $0.9$.
The quantities $a_{max}$ is the maximum values of $a_{i+1/2}$,
for all `$i$'.

In the section \ref{sec:tests}, we present various numerical tests to validate the implementation of the scheme. We also confirm the fourth-order accuracy of the method for nonlinear electron-plasma oscillations.

\begin{figure}
\vspace{-1.3in}
\includegraphics[width=1.0\columnwidth]{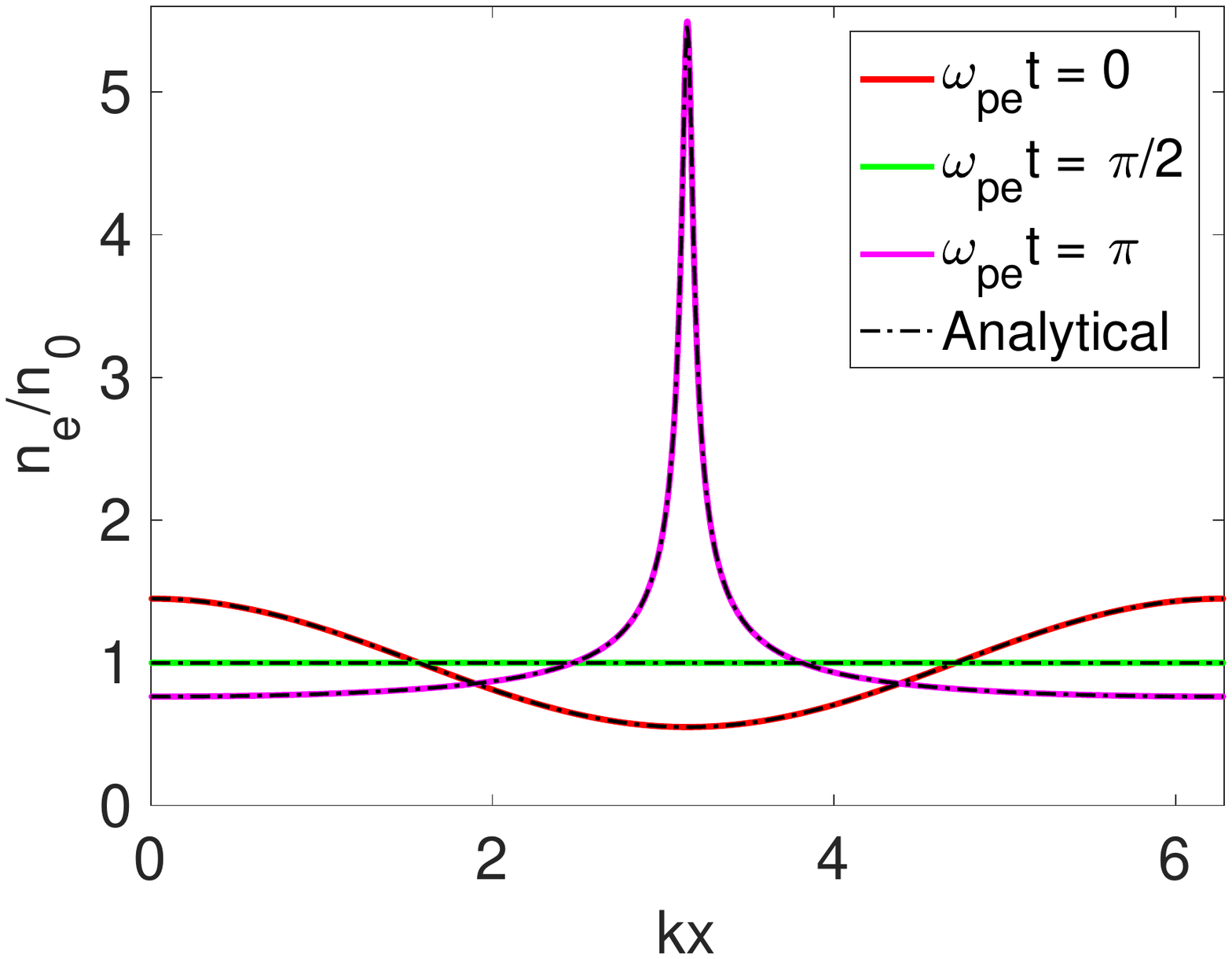}
\vspace{-1.4in}
\caption{\label{fig:fig1} Comparison between analytical (dashed line points) and numerical profiles (solid lines) of electron density for the nonlinear electron plasma oscillations in a cold non-dissipative plasma at an initial amplitude $\delta = 0.45$}
\end{figure}

\begin{figure}
\vspace{-1.3in}
\includegraphics[width=1.0\columnwidth]{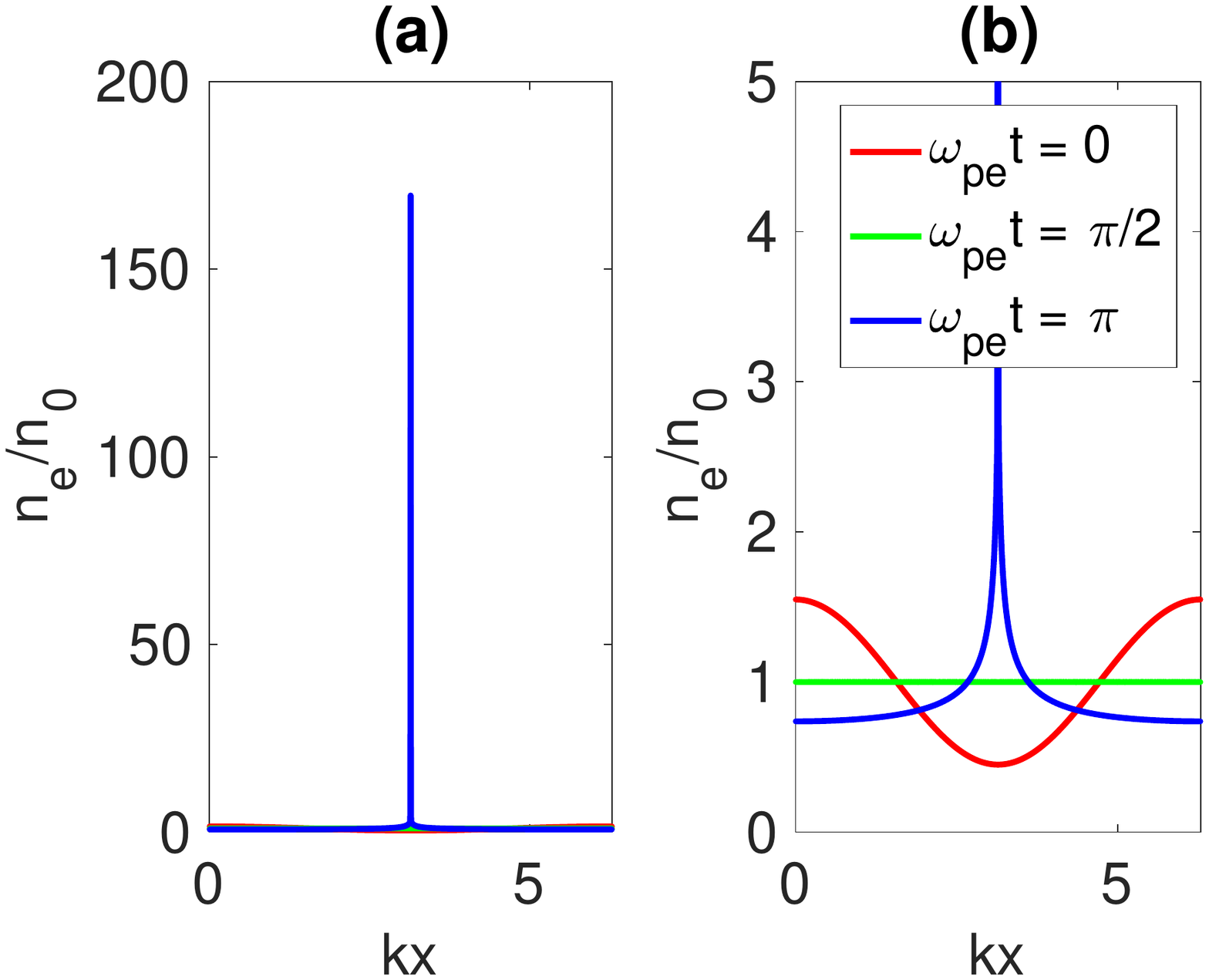}
\vspace{-1.5in}
\caption{ \label{fig:fig2} Space-time evolution of the electron density in the nonlinear electron plasma oscillations in a cold non-dissipative plasma at an initial amplitude $\delta = 0.55$.
The left figure (a) shows the density burst, and the right figure (b) is the zoomed version of the left figure which allows one to visualize the density profiles at $\omega_{pe}t=0$ (in red color) and $\omega_{pe}t=\pi/2.$ (in green color). The solid blue lines in both the sub-plots denote the density profile at $\omega_{pe}t=\pi$.  }
\end{figure}

\section{Numerical tests} \label{sec:tests}
\begin{figure}
\vspace{-1.3in}
\includegraphics[width=1.0\columnwidth]{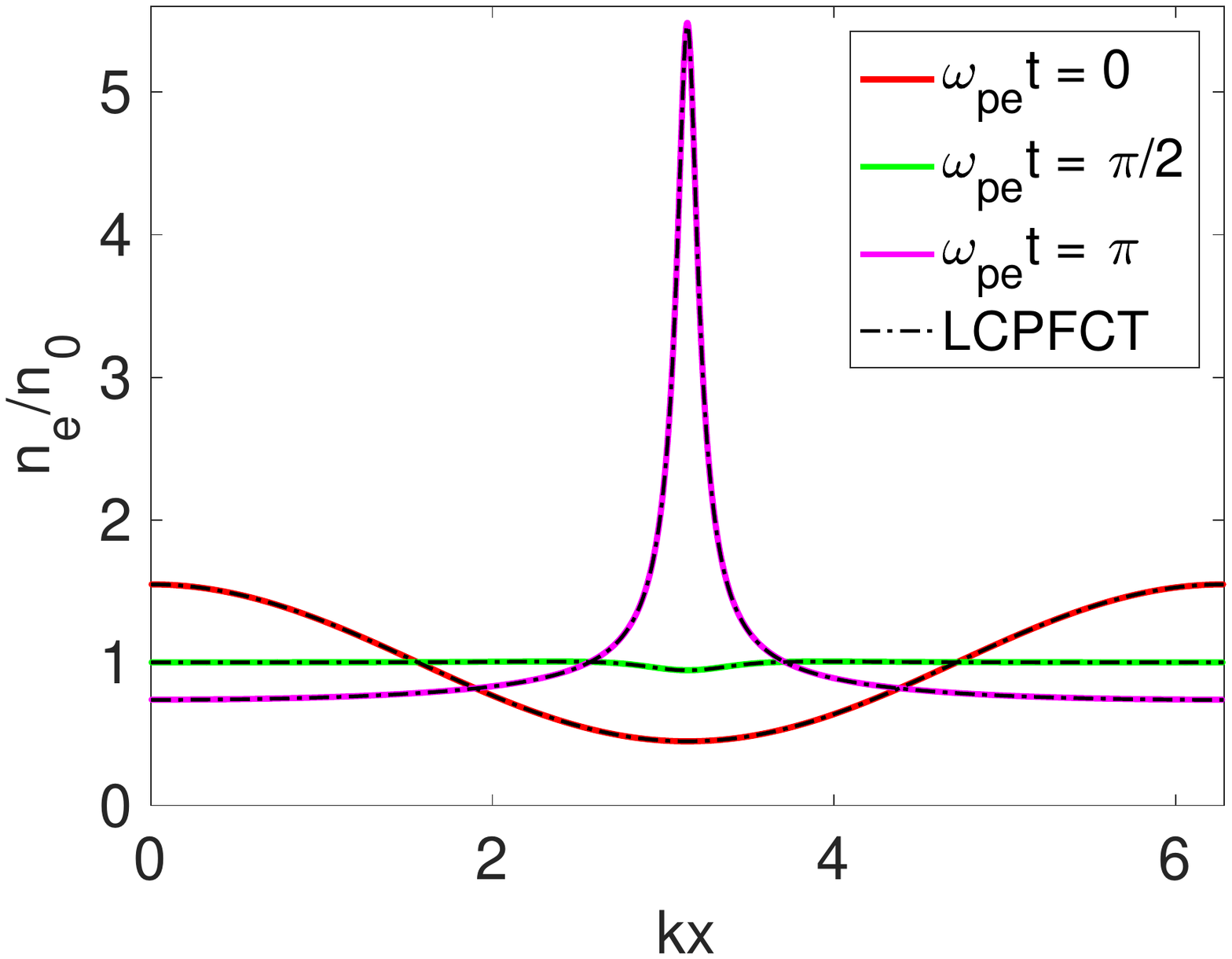}
\vspace{-1.3in}
\caption{ \label{fig:fig3}Comparison between our previous results from the LCPFCT (dashed line points) and from the present
numerical scheme (solid lines) where the profiles of electron density for the nonlinear electron plasma oscillations in a cold dissipative plasma at an initial amplitude $\delta = 0.45$ are shown.}
\end{figure}

\begin{figure}
\vspace{-1.2in}
\includegraphics[width=1.0\columnwidth]{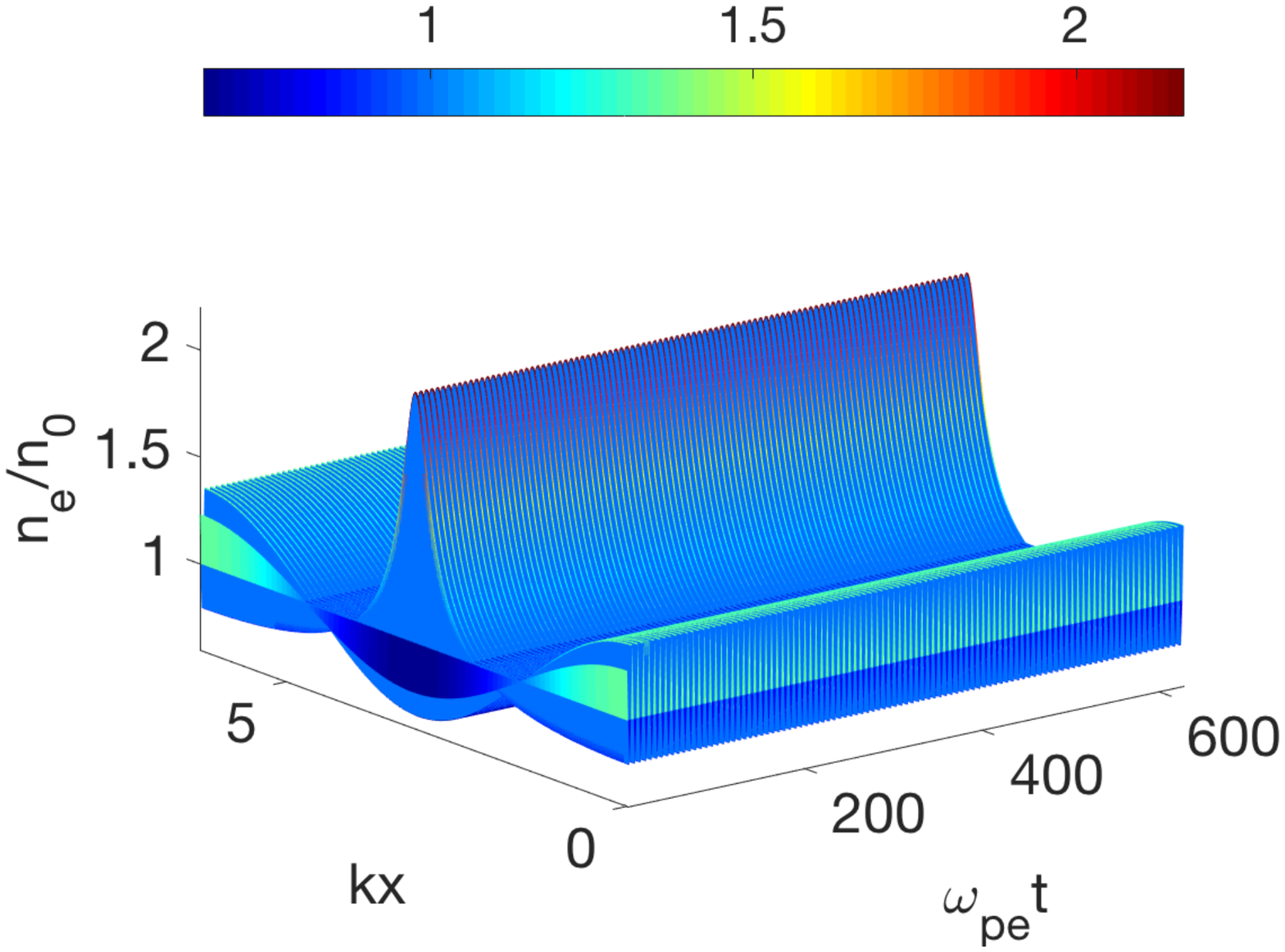}
\vspace{-1.5in}
\caption{ \label{fig:fig4} Space-time evolution of the electron density in a nonlinear electron plasma oscillations in a cold
non-dissipative plasma up to $\omega_{pe}t = 200\pi$ at an amplitude $\delta = 0.35$}
\end{figure}

\begin{table*}
\begin{minipage}{145mm}
\centering
\captionof{table}{Convergence of error and the EOC for the nonlinear electron plasma oscillations in a cold non-dissipative plasma
performed after one period at an amplitude $\delta = 0.35$.}
\label{tab:accu}
\begin{tabular}{ |cc|c|c|c|c|c|c|c| }
 \hline
\multicolumn{2}{|c|}{$j$} & $1$ & $2$ & $3$ & $4$ & $5$ & $6$ & $7$ \\
 \hline
\multicolumn{2}{|c|}{$n_x$} & $128$ & $256$ & $512$ & $1024$ & $2048$ & $4096$ & $8192$\\
\hline
\multirow{2}{*}{$L_1$} & $\delta U$ & 4.264 $10^{-4}$ & 2.761 $10^{-5}$ & 1.077 $10^{-6}$ & 3.622 $10^{-8}$ & 1.209 $10^{-9}$&
4.277 $10^{-11}$ & 2.133 $10^{-12}$\\
                        & EOC      & -               & 3.94             & 4.67 & 4.89 & 4.90 & 4.82 & 4.32\\
\hline
\end{tabular}
\end{minipage}
\end{table*}
For all the tests presented in this section we solve Eq.\eqref{conservative} using the periodic boundary conditions and the initial conditions are chosen as follows,
\begin{equation}
n_e(x,0) = 1 + \delta \cos x,\hspace{0.05cm} v_e(x,0) = 0,\hspace{0.05cm} E(x,0) = -\delta \sin x.     
\end{equation}
Number of grid points $n_x = 2048$, which remains the same in all the tests, excluding the convergence study. For the non-dissipative plasma $\nu_0 = \eta = 0$, however, for the dissipative plasma $\nu_0 = 0.03$ and $\eta = 2 \times 10^{-5}$ which are same as the ones chosen in Ref. \cite{verma2010nonlinear} for
one of the tests. Various values of $\delta$ are chosen for different tests. For example, in order to verify the correctness of the numerical scheme, in Fig.\ref{fig:fig1} we first compare the evolution of the nonlinear electron plasma oscillations in a cold non-dissipative with the previously established analytical results \cite{davidson1968nonlinear}. Here we choose $\delta = 0.45$ same as in \cite{davidson1968nonlinear}. The solid lines in the figure denote the results from our simulations, and the dashed line points stand for the analytical results \cite{davidson1968nonlinear}. A good agreement between the theory and the simulation is witnessed in this test.

It is well known \cite{davidson1968nonlinear} and also confirmed in numerical simulations
\cite{verma2011nonlinearkaw,verma2012residual} that in a cold non-dissipative plasma the nonlinear electron plasma oscillations, initiated by
a sinusoidal density perturbation, break at $\omega_{pe} t = \pi$ when the initial amplitude $\delta \ge 0.5$.
In Fig.\ref{fig:fig2} we show the evolution of the same at an initial amplitude $\delta = 0.55$ and observe a density burst
(a signature of wave-breaking) $\omega_{pe} t = \pi$. We stop the simulation at the first density burst because the results from the
fluid simulations are not physically valid after the wave-breaking.

In our previous work, we have numerically studied the nonlinear electron plasma oscillations in cold dissipative plasma for the
constant viscosity $\nu_0 = 0.03$ and resistivity $\eta = 2 \times 10^{-5}$ and found that the oscillations initiated by
the sinusoidal density perturbation do not break even when $\delta = 0.55 (> 0.5)$.
In Fig.\ref{fig:fig3}, we compare the results from the present and previous simulations for the same parameters, where
the solid lines in the figure are from the present simulation and the dashed line points represent the old results from the
LCPFCT simulation \cite{verma2010nonlinear}. Both the results exhibit a good agreement between the two.

In order to demonstrate the stability and the negligible numerical dissipation of the present numerical method, 
we show in Fig.\ref{fig:fig4} the space-time evolution of the electron density profile in a cold non-dissipative plasma up 
to one hundred electron plasma oscillations at an amplitude $\delta = 0.35$. No unphysical effects such as numerical noise are observed here.

Moreover, we also confirm the fourth-order accuracy of the numerical scheme, as shown in Table.\ref{tab:accu}, by estimating the error for the nonlinear electron plasma oscillations in a cold non-dissipative plasma at different spatial resolutions after one period $\omega_{pe} t = 2\pi$. He we compute the $L_1$ norm of the error as follows,

\begin{equation}
 \delta U = \frac{1}{n_x}\sum_{i = 1}^{n_x} |U_i^f - U_i^0|,  
\end{equation}
where $U_i^0$, $U_i^f$ are the initial (at $\omega_{pe} t = 0$) and the final (at $\omega_{pe} t = 2\pi$)
profiles of the electron density the numerical solutions as a function of grid resolution and
and $n_x$ is the number of grid points.

After computing the norms of the errors, we obtain the experimental order
of convergence $(EOC)$ using the formula,
\begin{equation}
  EOC(j+1) = \frac{|log(L_1(n_x(j+1)))|-|log(L_1(n_x(j)))|}{|log(n_x(j+1))|-|log(n_x(j))|},   
\end{equation}
here $j$ runs from $1$ to $7$. 

{\section{Summary and discussion}}
In summary, we have presented here an easy-to-implement higher-order finite volume scheme to study the nonlinear behavior of the non-relativistic 
electron-plasma oscillations in the cold dissipative and non-dissipative plasmas. The correctness of the new scheme has been established 
by reproducing previous analytical and numerical results for both non-dissipative and dissipative cold plasmas. The stability 
and the least dissipative nature of the numerical scheme have been confirmed by following the evolution of the nonlinear electron plasma oscillations up to hundred 
plasma periods, and
the fourth-order accuracy has also been demonstrated by evaluating the error at various grid resolutions after one plasma period.
Extension of the present method for the warm and magnetized plasma is straightforward. Implementations of the relativistic effects and
multi-component plasma are in progress and would be published elsewhere. 

\section*{Acknowledgments} The author would like to thank Dr. Yannick Marandet for valuable suggestions.


\begin{thebibliography}{40}%
\makeatletter
\providecommand \@ifxundefined [1]{%
 \@ifx{#1\undefined}
}%
\providecommand \@ifnum [1]{%
 \ifnum #1\expandafter \@firstoftwo
 \else \expandafter \@secondoftwo
 \fi
}%
\providecommand \@ifx [1]{%
 \ifx #1\expandafter \@firstoftwo
 \else \expandafter \@secondoftwo
 \fi
}%
\providecommand \natexlab [1]{#1}%
\providecommand \enquote  [1]{``#1''}%
\providecommand \bibnamefont  [1]{#1}%
\providecommand \bibfnamefont [1]{#1}%
\providecommand \citenamefont [1]{#1}%
\providecommand \href@noop [0]{\@secondoftwo}%
\providecommand \href [0]{\begingroup \@sanitize@url \@href}%
\providecommand \@href[1]{\@@startlink{#1}\@@href}%
\providecommand \@@href[1]{\endgroup#1\@@endlink}%
\providecommand \@sanitize@url [0]{\catcode `\\12\catcode `\$12\catcode
  `\&12\catcode `\#12\catcode `\^12\catcode `\_12\catcode `\%12\relax}%
\providecommand \@@startlink[1]{}%
\providecommand \@@endlink[0]{}%
\providecommand \url  [0]{\begingroup\@sanitize@url \@url }%
\providecommand \@url [1]{\endgroup\@href {#1}{\urlprefix }}%
\providecommand \urlprefix  [0]{URL }%
\providecommand \Eprint [0]{\href }%
\providecommand \doibase [0]{http://dx.doi.org/}%
\providecommand \selectlanguage [0]{\@gobble}%
\providecommand \bibinfo  [0]{\@secondoftwo}%
\providecommand \bibfield  [0]{\@secondoftwo}%
\providecommand \translation [1]{[#1]}%
\providecommand \BibitemOpen [0]{}%
\providecommand \bibitemStop [0]{}%
\providecommand \bibitemNoStop [0]{.\EOS\space}%
\providecommand \EOS [0]{\spacefactor3000\relax}%
\providecommand \BibitemShut  [1]{\csname bibitem#1\endcsname}%
\let\auto@bib@innerbib\@empty
\bibitem [{\citenamefont {Jain}\ \emph {et~al.}(2003)\citenamefont {Jain},
  \citenamefont {Das}, \citenamefont {Kaw},\ and\ \citenamefont
  {Sengupta}}]{jain2003nonlinear}%
  \BibitemOpen
  \bibfield  {author} {\bibinfo {author} {\bibfnamefont {N.}~\bibnamefont
  {Jain}}, \bibinfo {author} {\bibfnamefont {A.}~\bibnamefont {Das}}, \bibinfo
  {author} {\bibfnamefont {P.}~\bibnamefont {Kaw}}, \ and\ \bibinfo {author}
  {\bibfnamefont {S.}~\bibnamefont {Sengupta}},\ }\href@noop {} {\bibfield
  {journal} {\bibinfo  {journal} {Physics of Plasmas}\ }\textbf {\bibinfo
  {volume} {10}},\ \bibinfo {pages} {29} (\bibinfo {year} {2003})}\BibitemShut
  {NoStop}%
\bibitem [{\citenamefont {Verma}\ \emph {et~al.}(2010)\citenamefont {Verma},
  \citenamefont {Soni}, \citenamefont {Sengupta},\ and\ \citenamefont
  {Kaw}}]{verma2010nonlinear}%
  \BibitemOpen
  \bibfield  {author} {\bibinfo {author} {\bibfnamefont {P.~S.}\ \bibnamefont
  {Verma}}, \bibinfo {author} {\bibfnamefont {J.}~\bibnamefont {Soni}},
  \bibinfo {author} {\bibfnamefont {S.}~\bibnamefont {Sengupta}}, \ and\
  \bibinfo {author} {\bibfnamefont {P.}~\bibnamefont {Kaw}},\ }\href@noop {}
  {\bibfield  {journal} {\bibinfo  {journal} {Physics of Plasmas}\ }\textbf
  {\bibinfo {volume} {17}},\ \bibinfo {pages} {044503} (\bibinfo {year}
  {2010})}\BibitemShut {NoStop}%
\bibitem [{\citenamefont {Verma}\ \emph
  {et~al.}(2012{\natexlab{a}})\citenamefont {Verma}, \citenamefont {Sengupta},\
  and\ \citenamefont {Kaw}}]{verma2012breaking}%
  \BibitemOpen
  \bibfield  {author} {\bibinfo {author} {\bibfnamefont {P.~S.}\ \bibnamefont
  {Verma}}, \bibinfo {author} {\bibfnamefont {S.}~\bibnamefont {Sengupta}}, \
  and\ \bibinfo {author} {\bibfnamefont {P.}~\bibnamefont {Kaw}},\ }\href@noop
  {} {\bibfield  {journal} {\bibinfo  {journal} {Physical Review Letters}\
  }\textbf {\bibinfo {volume} {108}},\ \bibinfo {pages} {125005} (\bibinfo
  {year} {2012}{\natexlab{a}})}\BibitemShut {NoStop}%
\bibitem [{\citenamefont {Verma}\ \emph
  {et~al.}(2012{\natexlab{b}})\citenamefont {Verma}, \citenamefont {Sengupta},\
  and\ \citenamefont {Kaw}}]{verma2012residual}%
  \BibitemOpen
  \bibfield  {author} {\bibinfo {author} {\bibfnamefont {P.~S.}\ \bibnamefont
  {Verma}}, \bibinfo {author} {\bibfnamefont {S.}~\bibnamefont {Sengupta}}, \
  and\ \bibinfo {author} {\bibfnamefont {P.}~\bibnamefont {Kaw}},\ }\href@noop
  {} {\bibfield  {journal} {\bibinfo  {journal} {Physical Review E}\ }\textbf
  {\bibinfo {volume} {86}},\ \bibinfo {pages} {016410} (\bibinfo {year}
  {2012}{\natexlab{b}})}\BibitemShut {NoStop}%
\bibitem [{\citenamefont {Verma}(2017)}]{verma2017generation}%
  \BibitemOpen
  \bibfield  {author} {\bibinfo {author} {\bibfnamefont {P.~S.}\ \bibnamefont
  {Verma}},\ }\href@noop {} {\bibfield  {journal} {\bibinfo  {journal} {Physics
  Letters A}\ } (\bibinfo {year} {2017})}\BibitemShut {NoStop}%
\bibitem [{\citenamefont {Verma}\ and\ \citenamefont
  {Adhyapak}(2017)}]{verma2017nonlinear}%
  \BibitemOpen
  \bibfield  {author} {\bibinfo {author} {\bibfnamefont {P.~S.}\ \bibnamefont
  {Verma}}\ and\ \bibinfo {author} {\bibfnamefont {T.~C.}\ \bibnamefont
  {Adhyapak}},\ }\href@noop {} {\bibfield  {journal} {\bibinfo  {journal}
  {Physics of Plasmas}\ }\textbf {\bibinfo {volume} {24}},\ \bibinfo {pages}
  {112112} (\bibinfo {year} {2017})}\BibitemShut {NoStop}%
\bibitem [{\citenamefont {Dawson}(1959)}]{dawson1959nonlinear}%
  \BibitemOpen
  \bibfield  {author} {\bibinfo {author} {\bibfnamefont {J.~M.}\ \bibnamefont
  {Dawson}},\ }\href@noop {} {\bibfield  {journal} {\bibinfo  {journal}
  {Physical Review}\ }\textbf {\bibinfo {volume} {113}},\ \bibinfo {pages}
  {383} (\bibinfo {year} {1959})}\BibitemShut {NoStop}%
\bibitem [{\citenamefont {Davidson}\ and\ \citenamefont
  {Schram}(1968)}]{davidson1968nonlinear}%
  \BibitemOpen
  \bibfield  {author} {\bibinfo {author} {\bibfnamefont {R.}~\bibnamefont
  {Davidson}}\ and\ \bibinfo {author} {\bibfnamefont {P.}~\bibnamefont
  {Schram}},\ }\href@noop {} {\bibfield  {journal} {\bibinfo  {journal}
  {Nuclear Fusion}\ }\textbf {\bibinfo {volume} {8}},\ \bibinfo {pages} {183}
  (\bibinfo {year} {1968})}\BibitemShut {NoStop}%
\bibitem [{\citenamefont {Kaw}\ \emph {et~al.}(1973)\citenamefont {Kaw},
  \citenamefont {Lin},\ and\ \citenamefont {Dawson}}]{kaw1973quasiresonant}%
  \BibitemOpen
  \bibfield  {author} {\bibinfo {author} {\bibfnamefont {P.~K.}\ \bibnamefont
  {Kaw}}, \bibinfo {author} {\bibfnamefont {A.}~\bibnamefont {Lin}}, \ and\
  \bibinfo {author} {\bibfnamefont {J.}~\bibnamefont {Dawson}},\ }\href@noop {}
  {\bibfield  {journal} {\bibinfo  {journal} {The Physics of Fluids}\ }\textbf
  {\bibinfo {volume} {16}},\ \bibinfo {pages} {1967} (\bibinfo {year}
  {1973})}\BibitemShut {NoStop}%
\bibitem [{\citenamefont {Stenflo}\ and\ \citenamefont
  {Gradov}(1998)}]{stenflo1998electron}%
  \BibitemOpen
  \bibfield  {author} {\bibinfo {author} {\bibfnamefont {L.}~\bibnamefont
  {Stenflo}}\ and\ \bibinfo {author} {\bibfnamefont {O.}~\bibnamefont
  {Gradov}},\ }\href@noop {} {\bibfield  {journal} {\bibinfo  {journal}
  {Physical Review E}\ }\textbf {\bibinfo {volume} {58}},\ \bibinfo {pages}
  {8044} (\bibinfo {year} {1998})}\BibitemShut {NoStop}%
\bibitem [{\citenamefont {Gupta}\ and\ \citenamefont
  {Kaw}(1999)}]{gupta1999phase}%
  \BibitemOpen
  \bibfield  {author} {\bibinfo {author} {\bibfnamefont {S.~S.}\ \bibnamefont
  {Gupta}}\ and\ \bibinfo {author} {\bibfnamefont {P.~K.}\ \bibnamefont
  {Kaw}},\ }\href@noop {} {\bibfield  {journal} {\bibinfo  {journal} {Physical
  Review Letters}\ }\textbf {\bibinfo {volume} {82}},\ \bibinfo {pages} {1867}
  (\bibinfo {year} {1999})}\BibitemShut {NoStop}%
\bibitem [{\citenamefont {Rowlands}\ \emph {et~al.}(2008)\citenamefont
  {Rowlands}, \citenamefont {Brodin},\ and\ \citenamefont
  {Stenflo}}]{rowlands2008exact}%
  \BibitemOpen
  \bibfield  {author} {\bibinfo {author} {\bibfnamefont {G.}~\bibnamefont
  {Rowlands}}, \bibinfo {author} {\bibfnamefont {G.}~\bibnamefont {Brodin}}, \
  and\ \bibinfo {author} {\bibfnamefont {L.}~\bibnamefont {Stenflo}},\
  }\href@noop {} {\bibfield  {journal} {\bibinfo  {journal} {Journal of Plasma
  Physics}\ }\textbf {\bibinfo {volume} {74}},\ \bibinfo {pages} {569}
  (\bibinfo {year} {2008})}\BibitemShut {NoStop}%
\bibitem [{\citenamefont {Sengupta}\ \emph {et~al.}(2009)\citenamefont
  {Sengupta}, \citenamefont {Saxena}, \citenamefont {Kaw}, \citenamefont
  {Sen},\ and\ \citenamefont {Das}}]{sengupta2009phase}%
  \BibitemOpen
  \bibfield  {author} {\bibinfo {author} {\bibfnamefont {S.}~\bibnamefont
  {Sengupta}}, \bibinfo {author} {\bibfnamefont {V.}~\bibnamefont {Saxena}},
  \bibinfo {author} {\bibfnamefont {P.~K.}\ \bibnamefont {Kaw}}, \bibinfo
  {author} {\bibfnamefont {A.}~\bibnamefont {Sen}}, \ and\ \bibinfo {author}
  {\bibfnamefont {A.}~\bibnamefont {Das}},\ }\href@noop {} {\bibfield
  {journal} {\bibinfo  {journal} {Physical Review E}\ }\textbf {\bibinfo
  {volume} {79}},\ \bibinfo {pages} {026404} (\bibinfo {year}
  {2009})}\BibitemShut {NoStop}%
\bibitem [{\citenamefont {Maity}\ \emph {et~al.}(2010)\citenamefont {Maity},
  \citenamefont {Chakrabarti},\ and\ \citenamefont
  {Sengupta}}]{maity2010nonlinear}%
  \BibitemOpen
  \bibfield  {author} {\bibinfo {author} {\bibfnamefont {C.}~\bibnamefont
  {Maity}}, \bibinfo {author} {\bibfnamefont {N.}~\bibnamefont {Chakrabarti}},
  \ and\ \bibinfo {author} {\bibfnamefont {S.}~\bibnamefont {Sengupta}},\
  }\href@noop {} {\bibfield  {journal} {\bibinfo  {journal} {Physics of
  Plasmas}\ }\textbf {\bibinfo {volume} {17}},\ \bibinfo {pages} {082306}
  (\bibinfo {year} {2010})}\BibitemShut {NoStop}%
\bibitem [{\citenamefont {Maity}\ \emph {et~al.}(2012)\citenamefont {Maity},
  \citenamefont {Chakrabarti},\ and\ \citenamefont
  {Sengupta}}]{maity2012breaking}%
  \BibitemOpen
  \bibfield  {author} {\bibinfo {author} {\bibfnamefont {C.}~\bibnamefont
  {Maity}}, \bibinfo {author} {\bibfnamefont {N.}~\bibnamefont {Chakrabarti}},
  \ and\ \bibinfo {author} {\bibfnamefont {S.}~\bibnamefont {Sengupta}},\
  }\href@noop {} {\bibfield  {journal} {\bibinfo  {journal} {Physical Review
  E}\ }\textbf {\bibinfo {volume} {86}},\ \bibinfo {pages} {016408} (\bibinfo
  {year} {2012})}\BibitemShut {NoStop}%
\bibitem [{\citenamefont {Maity}\ \emph {et~al.}(2013)\citenamefont {Maity},
  \citenamefont {Sarkar}, \citenamefont {Shukla},\ and\ \citenamefont
  {Chakrabarti}}]{maity2013wave}%
  \BibitemOpen
  \bibfield  {author} {\bibinfo {author} {\bibfnamefont {C.}~\bibnamefont
  {Maity}}, \bibinfo {author} {\bibfnamefont {A.}~\bibnamefont {Sarkar}},
  \bibinfo {author} {\bibfnamefont {P.~K.}\ \bibnamefont {Shukla}}, \ and\
  \bibinfo {author} {\bibfnamefont {N.}~\bibnamefont {Chakrabarti}},\
  }\href@noop {} {\bibfield  {journal} {\bibinfo  {journal} {Physical Review
  Letters}\ }\textbf {\bibinfo {volume} {110}},\ \bibinfo {pages} {215002}
  (\bibinfo {year} {2013})}\BibitemShut {NoStop}%
\bibitem [{\citenamefont {Brodin}\ and\ \citenamefont
  {Stenflo}(2014{\natexlab{a}})}]{brodin2014large}%
  \BibitemOpen
  \bibfield  {author} {\bibinfo {author} {\bibfnamefont {G.}~\bibnamefont
  {Brodin}}\ and\ \bibinfo {author} {\bibfnamefont {L.}~\bibnamefont
  {Stenflo}},\ }\href@noop {} {\bibfield  {journal} {\bibinfo  {journal}
  {Physics Letters A}\ }\textbf {\bibinfo {volume} {378}},\ \bibinfo {pages}
  {1632} (\bibinfo {year} {2014}{\natexlab{a}})}\BibitemShut {NoStop}%
\bibitem [{\citenamefont {Brodin}\ and\ \citenamefont
  {Stenflo}(2014{\natexlab{b}})}]{brodin2014nonlinear}%
  \BibitemOpen
  \bibfield  {author} {\bibinfo {author} {\bibfnamefont {G.}~\bibnamefont
  {Brodin}}\ and\ \bibinfo {author} {\bibfnamefont {L.}~\bibnamefont
  {Stenflo}},\ }\href@noop {} {\bibfield  {journal} {\bibinfo  {journal}
  {Physics of Plasmas}\ }\textbf {\bibinfo {volume} {21}},\ \bibinfo {pages}
  {122301} (\bibinfo {year} {2014}{\natexlab{b}})}\BibitemShut {NoStop}%
\bibitem [{\citenamefont {Brodin}\ and\ \citenamefont
  {Stenflo}(2017{\natexlab{a}})}]{brodin2017simple}%
  \BibitemOpen
  \bibfield  {author} {\bibinfo {author} {\bibfnamefont {G.}~\bibnamefont
  {Brodin}}\ and\ \bibinfo {author} {\bibfnamefont {L.}~\bibnamefont
  {Stenflo}},\ }\href@noop {} {\bibfield  {journal} {\bibinfo  {journal}
  {Physics Letters A}\ }\textbf {\bibinfo {volume} {381}},\ \bibinfo {pages}
  {1033} (\bibinfo {year} {2017}{\natexlab{a}})}\BibitemShut {NoStop}%
\bibitem [{\citenamefont {Brodin}\ and\ \citenamefont
  {Stenflo}(2017{\natexlab{b}})}]{brodin2017nonlinear}%
  \BibitemOpen
  \bibfield  {author} {\bibinfo {author} {\bibfnamefont {G.}~\bibnamefont
  {Brodin}}\ and\ \bibinfo {author} {\bibfnamefont {L.}~\bibnamefont
  {Stenflo}},\ }\href@noop {} {\bibfield  {journal} {\bibinfo  {journal}
  {Physics of Plasmas}\ }\textbf {\bibinfo {volume} {24}},\ \bibinfo {pages}
  {124505} (\bibinfo {year} {2017}{\natexlab{b}})}\BibitemShut {NoStop}%
\bibitem [{\citenamefont {Faure}\ \emph {et~al.}(2004)\citenamefont {Faure},
  \citenamefont {Glinec}, \citenamefont {Pukhov}, \citenamefont {Kiselev},
  \citenamefont {Gordienko}, \citenamefont {Lefebvre}, \citenamefont
  {Rousseau}, \citenamefont {Burgy},\ and\ \citenamefont
  {Malka}}]{faure2004laser}%
  \BibitemOpen
  \bibfield  {author} {\bibinfo {author} {\bibfnamefont {J.}~\bibnamefont
  {Faure}}, \bibinfo {author} {\bibfnamefont {Y.}~\bibnamefont {Glinec}},
  \bibinfo {author} {\bibfnamefont {A.}~\bibnamefont {Pukhov}}, \bibinfo
  {author} {\bibfnamefont {S.}~\bibnamefont {Kiselev}}, \bibinfo {author}
  {\bibfnamefont {S.}~\bibnamefont {Gordienko}}, \bibinfo {author}
  {\bibfnamefont {E.}~\bibnamefont {Lefebvre}}, \bibinfo {author}
  {\bibfnamefont {J.-P.}\ \bibnamefont {Rousseau}}, \bibinfo {author}
  {\bibfnamefont {F.}~\bibnamefont {Burgy}}, \ and\ \bibinfo {author}
  {\bibfnamefont {V.}~\bibnamefont {Malka}},\ }\href@noop {} {\bibfield
  {journal} {\bibinfo  {journal} {Nature}\ }\textbf {\bibinfo {volume} {431}},\
  \bibinfo {pages} {541} (\bibinfo {year} {2004})}\BibitemShut {NoStop}%
\bibitem [{\citenamefont {Jain}\ \emph {et~al.}(2015)\citenamefont {Jain},
  \citenamefont {Antonsen~Jr},\ and\ \citenamefont
  {Palastro}}]{jain2015positron}%
  \BibitemOpen
  \bibfield  {author} {\bibinfo {author} {\bibfnamefont {N.}~\bibnamefont
  {Jain}}, \bibinfo {author} {\bibfnamefont {T.}~\bibnamefont {Antonsen~Jr}}, \
  and\ \bibinfo {author} {\bibfnamefont {J.}~\bibnamefont {Palastro}},\
  }\href@noop {} {\bibfield  {journal} {\bibinfo  {journal} {Physical Review
  Letters}\ }\textbf {\bibinfo {volume} {115}},\ \bibinfo {pages} {195001}
  (\bibinfo {year} {2015})}\BibitemShut {NoStop}%
\bibitem [{\citenamefont {Birdsall}\ and\ \citenamefont
  {Langdon}(2004)}]{birdsall2004plasma}%
  \BibitemOpen
  \bibfield  {author} {\bibinfo {author} {\bibfnamefont {C.~K.}\ \bibnamefont
  {Birdsall}}\ and\ \bibinfo {author} {\bibfnamefont {A.~B.}\ \bibnamefont
  {Langdon}},\ }\href@noop {} {\emph {\bibinfo {title} {Plasma physics via
  computer simulation}}}\ (\bibinfo  {publisher} {CRC Press},\ \bibinfo {year}
  {2004})\BibitemShut {NoStop}%
\bibitem [{\citenamefont {Mandal}\ and\ \citenamefont
  {Sharma}(2016)}]{mandal2016study}%
  \BibitemOpen
  \bibfield  {author} {\bibinfo {author} {\bibfnamefont {D.}~\bibnamefont
  {Mandal}}\ and\ \bibinfo {author} {\bibfnamefont {D.}~\bibnamefont
  {Sharma}},\ }in\ \href@noop {} {\emph {\bibinfo {booktitle} {Journal of
  Physics: Conference Series}}},\ Vol.\ \bibinfo {volume} {759}\ (\bibinfo
  {organization} {IOP Publishing},\ \bibinfo {year} {2016})\ p.\ \bibinfo
  {pages} {012068}\BibitemShut {NoStop}%
\bibitem [{\citenamefont {Schamel}\ \emph {et~al.}(2017)\citenamefont
  {Schamel}, \citenamefont {Mandal},\ and\ \citenamefont
  {Sharma}}]{schamel2017nonlinear}%
  \BibitemOpen
  \bibfield  {author} {\bibinfo {author} {\bibfnamefont {H.}~\bibnamefont
  {Schamel}}, \bibinfo {author} {\bibfnamefont {D.}~\bibnamefont {Mandal}}, \
  and\ \bibinfo {author} {\bibfnamefont {D.}~\bibnamefont {Sharma}},\
  }\href@noop {} {\bibfield  {journal} {\bibinfo  {journal} {Physics of
  Plasmas}\ }\textbf {\bibinfo {volume} {24}},\ \bibinfo {pages} {032109}
  (\bibinfo {year} {2017})}\BibitemShut {NoStop}%
\bibitem [{\citenamefont {Bera}\ \emph {et~al.}(2015)\citenamefont {Bera},
  \citenamefont {Sengupta},\ and\ \citenamefont {Das}}]{bera2015fluid}%
  \BibitemOpen
  \bibfield  {author} {\bibinfo {author} {\bibfnamefont {R.~K.}\ \bibnamefont
  {Bera}}, \bibinfo {author} {\bibfnamefont {S.}~\bibnamefont {Sengupta}}, \
  and\ \bibinfo {author} {\bibfnamefont {A.}~\bibnamefont {Das}},\ }\href@noop
  {} {\bibfield  {journal} {\bibinfo  {journal} {Physics of Plasmas}\ }\textbf
  {\bibinfo {volume} {22}},\ \bibinfo {pages} {073109} (\bibinfo {year}
  {2015})}\BibitemShut {NoStop}%
\bibitem [{\citenamefont {Bera}\ \emph {et~al.}(2016)\citenamefont {Bera},
  \citenamefont {Mukherjee}, \citenamefont {Sengupta},\ and\ \citenamefont
  {Das}}]{bera2016relativistic}%
  \BibitemOpen
  \bibfield  {author} {\bibinfo {author} {\bibfnamefont {R.~K.}\ \bibnamefont
  {Bera}}, \bibinfo {author} {\bibfnamefont {A.}~\bibnamefont {Mukherjee}},
  \bibinfo {author} {\bibfnamefont {S.}~\bibnamefont {Sengupta}}, \ and\
  \bibinfo {author} {\bibfnamefont {A.}~\bibnamefont {Das}},\ }\href@noop {}
  {\bibfield  {journal} {\bibinfo  {journal} {Physics of Plasmas}\ }\textbf
  {\bibinfo {volume} {23}},\ \bibinfo {pages} {083113} (\bibinfo {year}
  {2016})}\BibitemShut {NoStop}%
\bibitem [{\citenamefont {Boris}\ and\ \citenamefont
  {Book}(1997)}]{boris1997flux}%
  \BibitemOpen
  \bibfield  {author} {\bibinfo {author} {\bibfnamefont {J.~P.}\ \bibnamefont
  {Boris}}\ and\ \bibinfo {author} {\bibfnamefont {D.~L.}\ \bibnamefont
  {Book}},\ }\href@noop {} {\bibfield  {journal} {\bibinfo  {journal} {Journal
  of computational physics}\ }\textbf {\bibinfo {volume} {135}},\ \bibinfo
  {pages} {172} (\bibinfo {year} {1997})}\BibitemShut {NoStop}%
\bibitem [{\citenamefont {Levy}\ \emph
  {et~al.}(2000{\natexlab{a}})\citenamefont {Levy}, \citenamefont {Puppo},\
  and\ \citenamefont {Russo}}]{lpr7}%
  \BibitemOpen
  \bibfield  {author} {\bibinfo {author} {\bibfnamefont {D.}~\bibnamefont
  {Levy}}, \bibinfo {author} {\bibfnamefont {G.}~\bibnamefont {Puppo}}, \ and\
  \bibinfo {author} {\bibfnamefont {G.}~\bibnamefont {Russo}},\ }\href@noop {}
  {\bibfield  {journal} {\bibinfo  {journal} {Applied Numerical Mathematics}\
  }\textbf {\bibinfo {volume} {33}},\ \bibinfo {pages} {407} (\bibinfo {year}
  {2000}{\natexlab{a}})}\BibitemShut {NoStop}%
\bibitem [{\citenamefont {Shu}(2009)}]{shu2009high}%
  \BibitemOpen
  \bibfield  {author} {\bibinfo {author} {\bibfnamefont {C.-W.}\ \bibnamefont
  {Shu}},\ }\href@noop {} {\bibfield  {journal} {\bibinfo  {journal} {SIAM
  Review}\ }\textbf {\bibinfo {volume} {51}},\ \bibinfo {pages} {82} (\bibinfo
  {year} {2009})}\BibitemShut {NoStop}%
\bibitem [{\citenamefont {Levy}()}]{lpr1}%
  \BibitemOpen
  \bibfield  {author} {\bibinfo {author} {\bibfnamefont {D.}~\bibnamefont
  {Levy}},\ }\href@noop {} {\bibfield  {journal} {\bibinfo  {journal}
  {Syst{\'e}m hyperboliques: Nouveaux sch{\'e}mas et nouvelles applications}\
  }\textbf {\bibinfo {volume} {1}},\ \bibinfo {pages} {489}}\BibitemShut
  {NoStop}%
\bibitem [{\citenamefont {Bianco}\ \emph {et~al.}(1999)\citenamefont {Bianco},
  \citenamefont {Puppo},\ and\ \citenamefont {Russo}}]{lpr2}%
  \BibitemOpen
  \bibfield  {author} {\bibinfo {author} {\bibfnamefont {F.}~\bibnamefont
  {Bianco}}, \bibinfo {author} {\bibfnamefont {G.}~\bibnamefont {Puppo}}, \
  and\ \bibinfo {author} {\bibfnamefont {G.}~\bibnamefont {Russo}},\
  }\href@noop {} {\bibfield  {journal} {\bibinfo  {journal} {SIAM Journal on
  Scientific Computing}\ }\textbf {\bibinfo {volume} {21}},\ \bibinfo {pages}
  {294} (\bibinfo {year} {1999})}\BibitemShut {NoStop}%
\bibitem [{\citenamefont {Levy}\ \emph {et~al.}(1999)\citenamefont {Levy},
  \citenamefont {Puppo},\ and\ \citenamefont {Russo}}]{lpr3}%
  \BibitemOpen
  \bibfield  {author} {\bibinfo {author} {\bibfnamefont {D.}~\bibnamefont
  {Levy}}, \bibinfo {author} {\bibfnamefont {G.}~\bibnamefont {Puppo}}, \ and\
  \bibinfo {author} {\bibfnamefont {G.}~\bibnamefont {Russo}},\ }\href@noop {}
  {\bibfield  {journal} {\bibinfo  {journal} {ESAIM: Mathematical Modelling and
  Numerical Analysis}\ }\textbf {\bibinfo {volume} {33}},\ \bibinfo {pages}
  {547} (\bibinfo {year} {1999})}\BibitemShut {NoStop}%
\bibitem [{\citenamefont {Levy}\ \emph {et~al.}(2002)\citenamefont {Levy},
  \citenamefont {Puppo},\ and\ \citenamefont {Russo}}]{lpr4}%
  \BibitemOpen
  \bibfield  {author} {\bibinfo {author} {\bibfnamefont {D.}~\bibnamefont
  {Levy}}, \bibinfo {author} {\bibfnamefont {G.}~\bibnamefont {Puppo}}, \ and\
  \bibinfo {author} {\bibfnamefont {G.}~\bibnamefont {Russo}},\ }\href@noop {}
  {\bibfield  {journal} {\bibinfo  {journal} {SIAM Journal on scientific
  computing}\ }\textbf {\bibinfo {volume} {24}},\ \bibinfo {pages} {480}
  (\bibinfo {year} {2002})}\BibitemShut {NoStop}%
\bibitem [{\citenamefont {Levy}\ \emph
  {et~al.}(2000{\natexlab{b}})\citenamefont {Levy}, \citenamefont {Puppo},\
  and\ \citenamefont {Russo}}]{lpr5}%
  \BibitemOpen
  \bibfield  {author} {\bibinfo {author} {\bibfnamefont {D.}~\bibnamefont
  {Levy}}, \bibinfo {author} {\bibfnamefont {G.}~\bibnamefont {Puppo}}, \ and\
  \bibinfo {author} {\bibfnamefont {G.}~\bibnamefont {Russo}},\ }\href@noop {}
  {\bibfield  {journal} {\bibinfo  {journal} {Applied Numerical Mathematics}\
  }\textbf {\bibinfo {volume} {33}},\ \bibinfo {pages} {415} (\bibinfo {year}
  {2000}{\natexlab{b}})}\BibitemShut {NoStop}%
\bibitem [{\citenamefont {Levy}\ \emph
  {et~al.}(2000{\natexlab{c}})\citenamefont {Levy}, \citenamefont {Puppo},\
  and\ \citenamefont {Russo}}]{lpr6}%
  \BibitemOpen
  \bibfield  {author} {\bibinfo {author} {\bibfnamefont {D.}~\bibnamefont
  {Levy}}, \bibinfo {author} {\bibfnamefont {G.}~\bibnamefont {Puppo}}, \ and\
  \bibinfo {author} {\bibfnamefont {G.}~\bibnamefont {Russo}},\ }\href@noop {}
  {\bibfield  {journal} {\bibinfo  {journal} {SIAM Journal on Scientific
  Computing}\ }\textbf {\bibinfo {volume} {22}},\ \bibinfo {pages} {656}
  (\bibinfo {year} {2000}{\natexlab{c}})}\BibitemShut {NoStop}%
\bibitem [{\citenamefont {Verma}\ and\ \citenamefont
  {M{\"u}ller}(2017)}]{verma2017higher}%
  \BibitemOpen
  \bibfield  {author} {\bibinfo {author} {\bibfnamefont {P.~S.}\ \bibnamefont
  {Verma}}\ and\ \bibinfo {author} {\bibfnamefont {W.-C.}\ \bibnamefont
  {M{\"u}ller}},\ }\href@noop {} {\bibfield  {journal} {\bibinfo  {journal}
  {Journal of Scientific Computing}\ ,\ \bibinfo {pages} {1}} (\bibinfo {year}
  {2017})}\BibitemShut {NoStop}%
\bibitem [{\citenamefont {Kurganov}\ and\ \citenamefont
  {Levy}(2000)}]{kurganov-2000}%
  \BibitemOpen
  \bibfield  {author} {\bibinfo {author} {\bibfnamefont {A.}~\bibnamefont
  {Kurganov}}\ and\ \bibinfo {author} {\bibfnamefont {D.}~\bibnamefont
  {Levy}},\ }\href@noop {} {\bibfield  {journal} {\bibinfo  {journal} {SIAM
  Journal on Scientific Computing}\ }\textbf {\bibinfo {volume} {22}},\
  \bibinfo {pages} {1461} (\bibinfo {year} {2000})}\BibitemShut {NoStop}%
\bibitem [{\citenamefont {Williamson}(1980)}]{rk4}%
  \BibitemOpen
  \bibfield  {author} {\bibinfo {author} {\bibfnamefont {J.}~\bibnamefont
  {Williamson}},\ }\href@noop {} {\bibfield  {journal} {\bibinfo  {journal}
  {Journal of Computational Physics}\ }\textbf {\bibinfo {volume} {35}},\
  \bibinfo {pages} {48} (\bibinfo {year} {1980})}\BibitemShut {NoStop}%
\bibitem [{\citenamefont {Verma}\ \emph {et~al.}(2011)\citenamefont {Verma},
  \citenamefont {Sengupta},\ and\ \citenamefont {Kaw}}]{verma2011nonlinearkaw}%
  \BibitemOpen
  \bibfield  {author} {\bibinfo {author} {\bibfnamefont {P.~S.}\ \bibnamefont
  {Verma}}, \bibinfo {author} {\bibfnamefont {S.}~\bibnamefont {Sengupta}}, \
  and\ \bibinfo {author} {\bibfnamefont {P.~K.}\ \bibnamefont {Kaw}},\
  }\href@noop {} {\bibfield  {journal} {\bibinfo  {journal} {Physics of
  Plasmas}\ }\textbf {\bibinfo {volume} {18}},\ \bibinfo {pages} {012301}
  (\bibinfo {year} {2011})}\BibitemShut {NoStop}%
\end{thebibliography}
%
\end{document}